\documentclass[english,reprint,aip,apl,twocoluomn,superscriptaddress,showpacs,amsmath]{revtex4-1}
\usepackage[T1]{fontenc}
\usepackage[latin9]{inputenc}
\usepackage{color}
\usepackage{units}
\usepackage{amssymb}
\usepackage{graphicx} %include graphicx
\usepackage{esint}
\usepackage{bm}     %bold match
\usepackage{natbib} %Bibilography with author year and number reference
\usepackage{ulem}  %for underlining
\usepackage{multirow}
\usepackage[colorlinks=true, citecolor=red]{hyperref}
\setcitestyle{journalcolor=blue}
\usepackage{babel}
\usepackage{dcolumn}
\begin{document}
%%%%%%%%%%Title and Author%%%%%%%%%%%
\title{Negative to Positive Magnetoresistance transition in Functionalization of Carbon nanotube and Polyaniline Composite}
\author{Krishna Prasad Maity}
\email{makrishna@iisc.ac.in}
\affiliation{Department of Physics, Indian Institute of Science, Bangalore 560012, India}
\author{Narendra Tanty , Ananya Patra, V Prasad}
\affiliation{Department of Physics, Indian Institute of Science, Bangalore 560012, India}
%\author{Narendra Tanty}
%\affiliation{Department of Physics, Indian Institute of Science, Bangalore 560012, India}
%\author{Ananya Patra}
%\affiliation{Department of Physics, Indian Institute of Science, Bangalore 560012, India}
%\author{V Prasad }
%\affiliation{Department of Physics, Indian Institute of Science, Bangalore 560012, India}

\begin{abstract}
Electrical resistivity and magnetoresistance(MR) in polyaniline(PANI) with carbon nanotube(CNT) and functionalized carbon nanotube(fCNT) composites have been studied for different weight percentages down to the temperature 4.2K and up to magnetic field 5T. Resistivity increases significantly in composite at low temperature due to functionalization of CNT compared to  only CNT. Interestingly transition from negative to positive magnetoresistance has been observed for 10wt\% of composite as the effect of disorder is more in fCNT/PANI. This result depicts that the MR has strong dependency on disorder in the composite system. The transition of MR has been explained on the basis of polaron-bipolaron model. The long range Coulomb interaction between two polarons screened by disorder in the composite of fCNT/PANI, increases the effective on-site Coulomb repulsion energy to form bipolaron which leads to change the sign of MR from negative to positive.  
\end{abstract}

\maketitle

\section{Introduction}
In recent years PANI with CNT and fCNT composites have been promising materials in application of gas sensing,\cite{zhang2008recent} DNA sensor, electromagnetic shielding(EMI), electrode of supercapacitor, printable circuit wiring, transparent covalent coating\cite{ramasubramaniam2003homogeneous,sahoo2010polymer}etc. Functionalization of CNT prevents the nanotube aggregation, improves interfacial interaction, dispersion and stabilization in polymer matrix. However, it shortens the length, breaks C-C $ sp^2$ bonds, enhances the disorder creating defects on sidewall and opening of the ends.~\cite{datsyuk2008chemical,rebelo2016progress,naseh2009functionalization} It is reported that the disorder in the system affects on the conductivity which can be changed several order of magnitude at low temperature. The conductivity of CNT/PANI composites change two order of magnitude by increasing the filler percentage due to strong coupling between CNT and lightly coated polymer chain.~\cite{long2004synthesis} The magnetoresistance study of doped PANI at high magnetic field and low temperature gives an idea of charge transport on local molecular level ordering in the system.~\cite{mukherjee2005magnetotransport} There is a transition from negative to positive MR, observed with lowering temperature, increasing magnetic field and decreasing CNT percentage in the composites.~\cite{long2008low} Also, Gu et. al has reported the transition from  positive to negative MR in disordered PANI-silicon nano composite around 5.5T at room temperature~\cite{gu2013separating} and explained by `Wavefunction shrinkage', `Forward interference' model. Adjusting the `dissociation', `charge reaction' mechanism of charge transport, MR has been tuned between positive and negative values.~\cite{hu2007tuning} The inversion of MR in organic semiconductor device has been shown depending on applied voltage, temperature and layer thickness.~\cite{bergeson2008inversion} Recently, the crossover of MR from positive to negative at 100K in doped polyaniline nanofibers has been reported and explained by `Bipolaron' model.~\cite{nath2013effect} Also Mamru et.al has studied that polyaniline composite shows MR transition at room temperature by varying the concentration of titania quantum dot due to suppression of polaron at functionalized PANI and titania interface.~\cite{mombru2017positive}
Motivated by these works we have studied the conductivity and MR of fCNT/PANI and CNT/PANI composites at low temperatures. To the best of our knowledge the transport properties of fCNT and composites with polymer has not been explored much. Hence this study will help us to understand the effect of disorder in charge transport mechanism of fCNT/PANI compare to CNT/PANI composite systems.  

\section{Methods}

We have synthesized polyaniline with multi-walled carbon nanotube(CNT) and functionalized carbon nanotube(fCNT) composites of different weight percentages(5,10 and 15wt$\%$) by in-situ chemical polymerization. Functionalization of CNT was done by immersing CNT in the mixture of concentrated $H_2SO_4$ and concentrated $HNO_3$  (3:1, volume ratio) for 24 hours at room temperature. Then fCNT was washed several times with DI water, filtered and dried in vacuum. We have followed the well known procedure to synthesize PANI composite with CNT and fCNT.~\cite{long2004synthesis,chakraborty2012effect,maser2003synthesis}
 In this process we added 2ml aniline monomer with 40ml 1.0M HCl, required amount of CNT/fCNT was added to make different weight percentage comoposites and sonicated for 1 hour to obtain well dispersed suspensions, then 400mg Ammonium per sulfate(APS) with 20ml 1.0 HCl solution was added dropwise into the above suspension at room temperature. Total suspension was stirred at 450rpm for 24 hours. After filtering, it was dried in vacuum oven for 48 hours at temperature $60^{0}C$. We collected the sample in powder form. Making pellet we have measured the resistivity and magnetoresistance by using Van der Pauw four probe method. We have done low temperature measurement using Janis liquid helium cryostat equipped with superconducting magnet.

\section{Characterization}
\subsection{Raman Spectroscopy}

\begin{figure}[tbh]
\begin{center}
\includegraphics[width=0.53\textwidth]{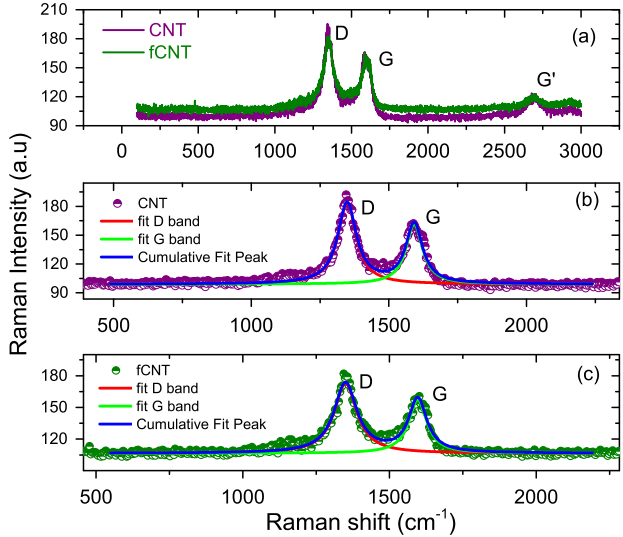}
\small{\caption{(a) Raman shift of CNT and fCNT. (b) and (c) D peak and G peak is fitted for CNT and fCNT respectively.\label{fig:raman}}}
\end{center}
\end{figure}

The effect of functionalization in CNT has been characterized by Raman spectroscopy. The Raman spectra of CNT,fCNT and their composites with PANI were excited by 514.5 nm laser. In figure~\ref{fig:raman} the Raman spectra of CNT and fCNT have been shown. CNT and fCNT consist of three characteristic bands, namely the D-band at 1348 $cm^{-1}$,the G-band at 1594 $cm^{-1}$, and the G'-band at 2692 $cm^{-1}$. The D-band and the G'-band are the disorder induced features arising from double resonance Raman scattering process from a non-zero center phonon mode.~\cite{ferrari2007raman,pimenta2007studying} The D-band generally attributed to the presence of amorphous or disorder carbon in the CNT sample.
\begin{figure}[tbh]
	\begin{center}
		\includegraphics[width=0.52\textwidth]{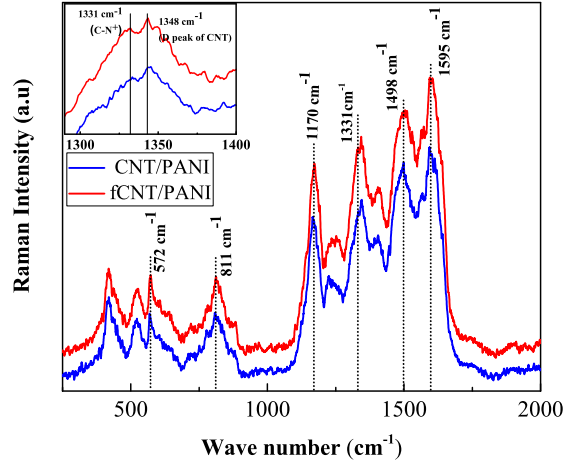}
		\small{\caption{Raman shift of CNT/PANI and fCNT/PANI composites.\label{fig:ramanf}}}
	\end{center}
\end{figure}

The integrated ratio $\frac{I_D}{I_G}$ has increased from 1.49 to 1.64
and FWHM of D-band has enhanced from 81.67 $cm^{-1}$ to 94.22 $cm^{-1}$ after functionalization of CNT. The Raman spectra of CNT/PANI and fCNT/PANI composites are shown in figure-\ref{fig:ramanf}. The characteristic peaks at 1348$cm^{-1}$(D band) and 1595$cm^{-1}$(G band) are also present in the composites.For both the composites C-N-C-H out of plane deformation at 572$cm^{-1}$, C-H wagging at 811$cm^{-1}$, C-H bending of quinoid ring at 1170 $cm^{-1}$, C-H bending of benzoid at 1348 $cm^{-1}$, C-N stretching at 1404 $cm^{-1}$ and C-C stretching of benzoid ring at 1498 $cm^{-1}$ were observed. The C-$N^{+\bullet}$ stretching peak at 1331$cm^{-1}$ represents polaron density.\cite{cochet2000theoretical} In the inset of fig~\ref{fig:ramanf} it is clearly shown that the intensity of the peak has increased significantly which implies the number of polaron has increased after functionalization of CNT.\cite{mombru2017positive}

\section{Result and Discussion}
\begin{figure}[tbh]
	\begin{center}
		\includegraphics[width=0.55\textwidth]{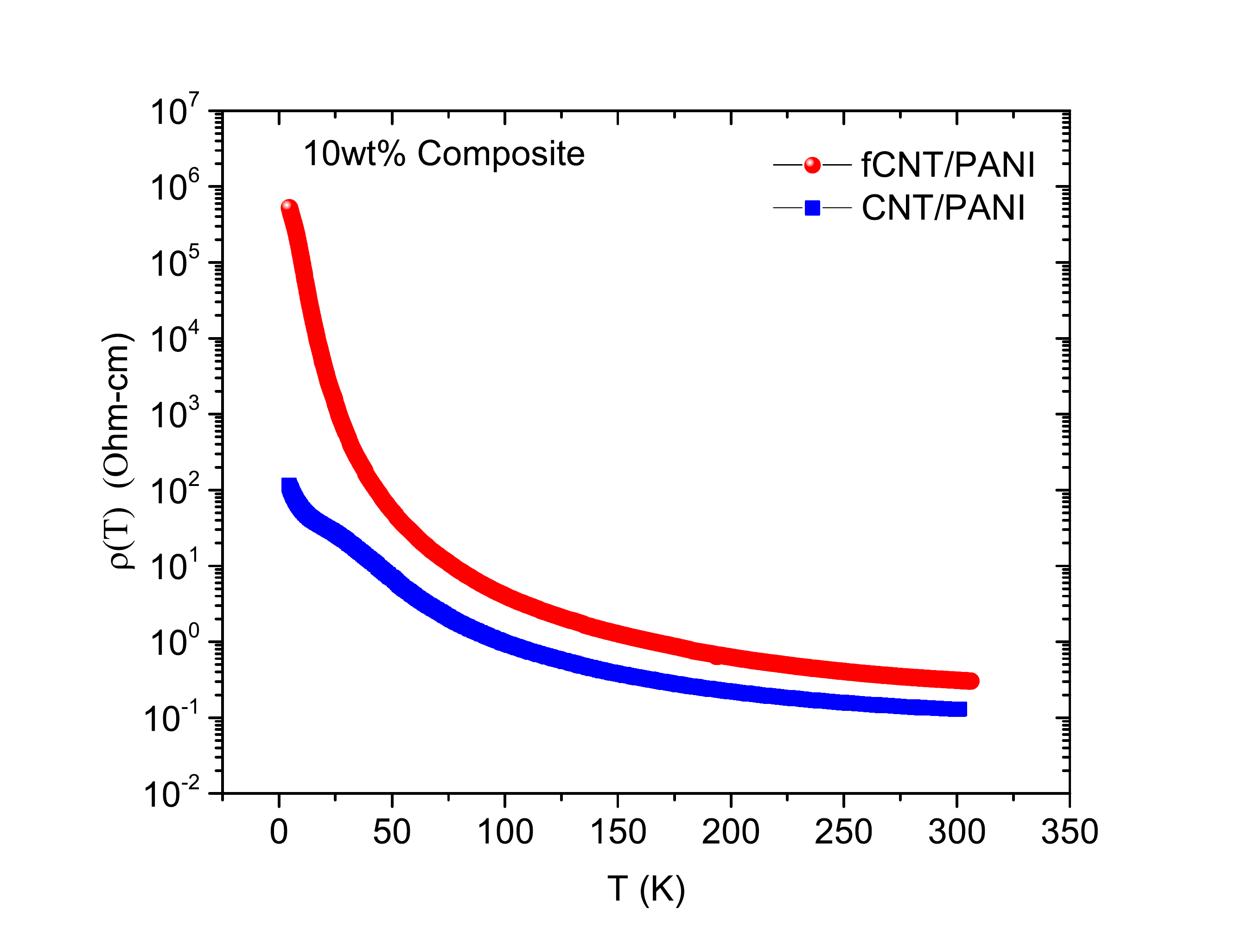}
		\small{\caption{(a) The temperature dependence resistivity for 10wt\% fCNT/PANI and CNT/PANI \label{fig:comrtagain}}}
	\end{center}
\end{figure}

\begin{figure}[tbh]
	\begin{center}
		\includegraphics[width=0.5\textwidth]{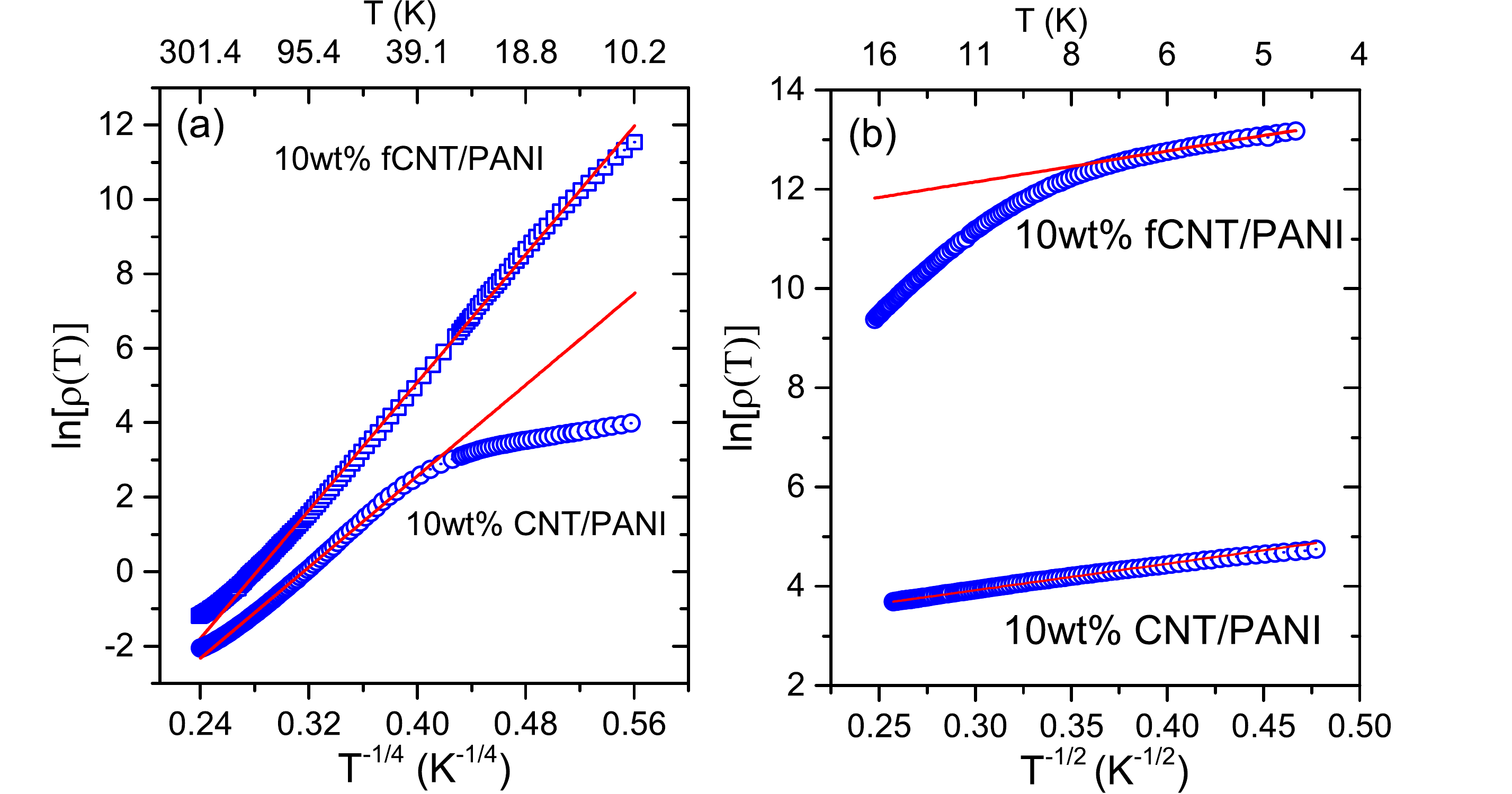}
		\small{\caption{(a) Plot of $ln\rho$ vs $T^{-1/4}$ above 15K fitted with Mott's 3D VRH model .(b) Plot $ln\rho$ vs $T^{-1/2}$ below 15K fitted with Efros and Shklovskii(ES) model for 10wt\% composite \label{fig:newrtcom}}}
	\end{center}
\end{figure}

The temperature dependence resistivity of CNT/PANI and fCNT/PANI for 10wt\% has been shown in figure~\ref{fig:comrtagain}. In both cases the resistivity increases significantly at low temperature from room temperature. The increase in resistivity for CNT/PANI composite has been observed before by several groups.~\cite{long2004synthesis,long2008low} We observed that for fCNT/PANI, resistivity increases by four order of magnitude more compared to CNT/PANI. The dependence of resistivity with the temperature has been explained by the Mott VRH model, which describes the low temperature resistivity in the strongly disordered system at the localized states.\cite{mott2012electronic} The model relates resistivity with temperature as 
\begin{eqnarray}
\rho(T)=\rho_0exp[(\frac{T_0}{T})^\frac{1}{d+1}]
\label{VRH}
\end{eqnarray}
 where $\rho_0$
is the resistivity at T=0K, d is the dimension of system, $k_B$ is the Boltzmann constant, T$_0$ is the Mott's temperature, 
$T_0=24/[\pi k_B N(E_F) a_0^3]$ which describes the energy barrier of an electron to hop from one localized state to the another. This energy barrier highly depends on the localization length and density of states at Fermi level(N($E_F$)). In figure~\ref{fig:newrtcom}(a) we have shown that 3D VRH [d=3] model is well fitted from room temperature to the temperature 35K and 15K for CNT/PANI and fCNT/PANI respectively. We have extracted the value of $T_0$ from  equation~\ref{VRH} above 15K as $8.78\times10^{5}K$ and $3.42\times10^{6}K$ for CNT/PANI and fCNT/PANI composite respectively. We have noticed that when CNT weight percentage increases in CNT/PANI composite the $T_0$ value decreases due to more pronounced effect of CNT in conduction. In the same weight percentage of fCNT/PANI composite, $T_0$ value is greater than CNT/PANI composite as fCNT contributes more disorder in the composite system.

Considering the coulomb interaction between the localized states which creates a small Coulomb gap $\Delta_{c}$ in the $N(E_F)$, Efros and Shklovskii(ES) showed that the $N(E_F)$ should quadratically vanish at the Fermi level~\cite{efros1975coulomb} and equation~\ref{VRH} is modified to
\begin{eqnarray}
\rho(T)=\rho_0 exp[({T_{ES}}/{T})^{1/2}]
\label{ES}
\end{eqnarray}
and the Coulomb gap
\begin{eqnarray}
\Delta_{c} = 0.905k_{B}T_{0}^{-1/2}T_{ES}^{3/2}
\label{ES2}
\end{eqnarray}
in all dimensions. where $T_{ES}$ is the Efros-Shklovskii temperature.
\begin{table}[tbh]
       \caption{Experimental values and VRH parameters for different samples, fCNT/PANI, CNT/PANI. The parameters are $T_{0}$, Mott characteristic temperature; $T_{ES}$, Efros-Shklovskii temperature; $\Delta_{c}$, Coulomb gap energy}.
		\resizebox{7cm}{!}{
	\begin{tabular}{cccc}
		\hline\hline
		sample & $T_{0}(K)$ & $T_{ES}(K)$ & $\Delta_{c}(meV)$ \\ [0.5ex]
		\hline  
		fCNT/PANI & $3.42\times 10^{6}$ & 38.8 &  0.010 \\ [1ex]
		CNT/PANI  & $8.78\times 10^{5}$ & 28.8 &  0.013 \\ [1ex]
		\hline\hline
	\end{tabular}
       }
       \label{table:1}
	\end{table}
\begin{figure}[tbh]
	\begin{center}
		\includegraphics[width=0.5\textwidth]{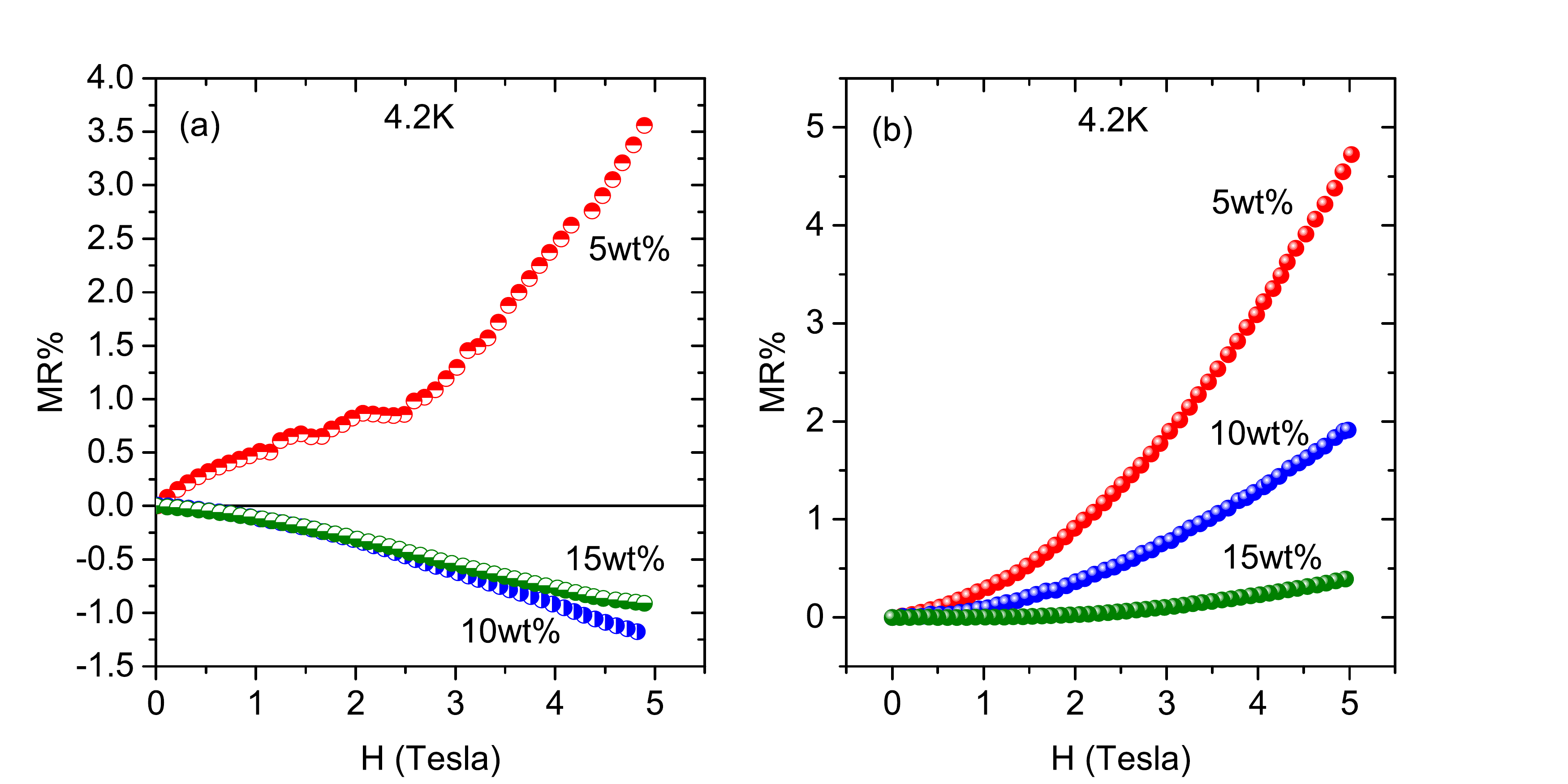}
		\small{\caption{(a) and (b) Plot of MR\% vs magnetic field for different wt\% of CNT/PANI and fCNT/PANI composites at 4.2K respectively.\label{fig:finalmr}}}
	\end{center}
\end{figure} 
 Figure~\ref{fig:newrtcom}(b) shows that resistivity data is well fitted with ES model below 10K and 15K for fCNT/PANI and CNT/PANI composite respectively. It depicts that the crossover from VRH to ES conduction took place from higher temperature for CNT/PANI composite compared to fCNT/PANI.The Coulomb energy gap decreases in fCNT compared to CNT composite(see Table~\ref{table:1}) which attributes the screening of the long-range coulomb interaction increases due to more disorder in the system. This result is consistent with previously reported data.\cite{ghosh1998transport,long2005electronic}  

\begin{figure}[tbh]
	\begin{center}
		\includegraphics[width=0.5\textwidth]{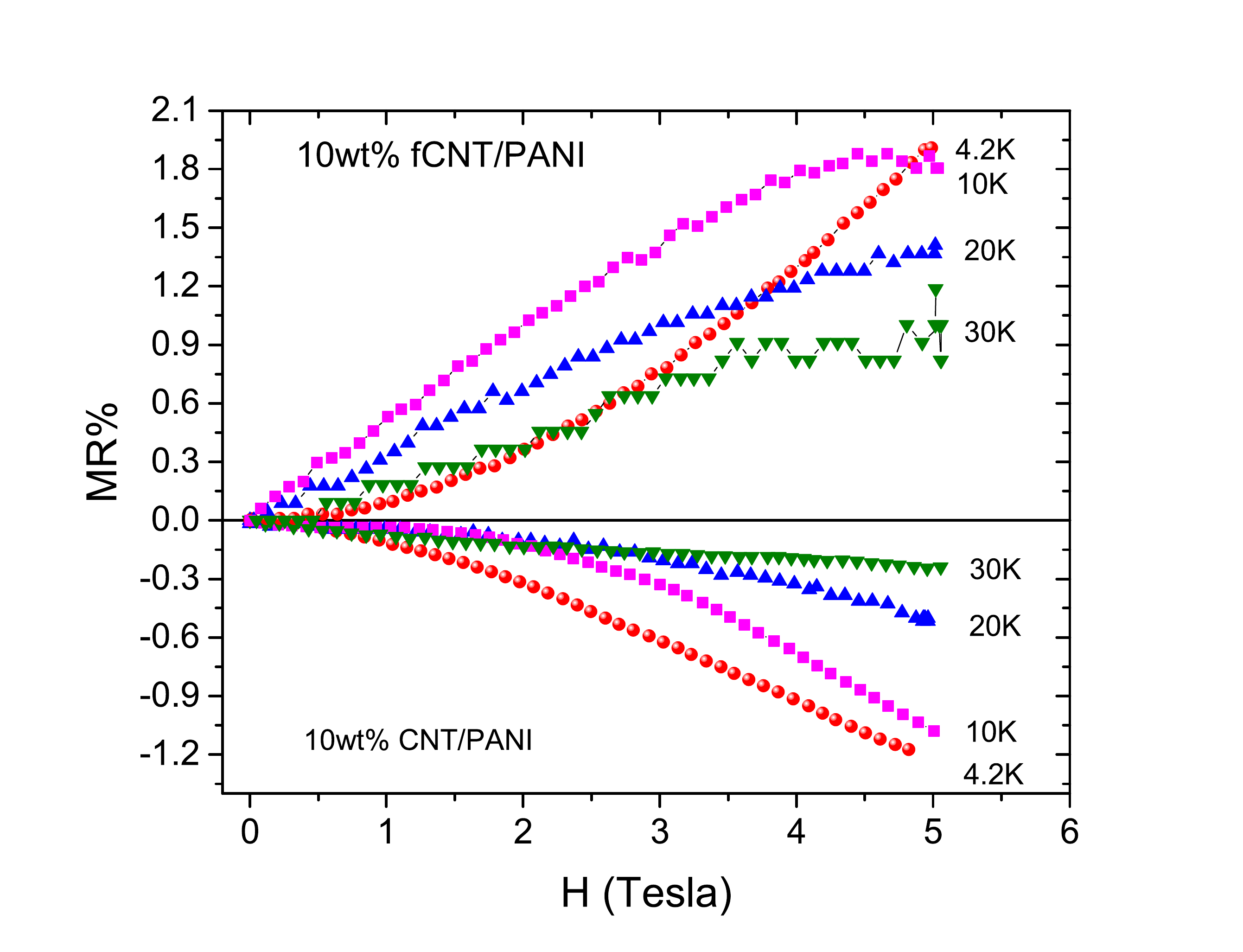}
		\small{\caption{ Plot of MR\% vs magnetic field for differnt temperature of 10wt\% CNT/PANI and fCNT/PANI.\label{fig:final10}}}
	\end{center}
\end{figure}

The variation of resistance with applied external magnetic field is known as  MR [ defined as [R(H)-R(0)]/R(0), in percentage]. Recent years in many systems like $Al_{70}Pd_{22.5}Re_{7.5}$~\cite{su2002magnetoresistance} and $Ni_xSi_{1-x}$~\cite{rosenbaum2001magnetoresistance} thin films have shown magnetoresistance changes sign from negative to positive by increasing magnetic field due to high disorder in the systems. In figure~\ref{fig:finalmr} we have plotted MR for composites of different weight percent of CNT and fCNT. It is observed that MR is always positive for all three wt\% of fCNT/PANI (ref. fig.~\ref{fig:finalmr}(b)) and there is a transition from positive to negative MR with increase in weight percentage of CNT from 5wt\% to 10wt\% in CNT/PANI(fig.~\ref{fig:finalmr}(a)). Interestingly the value of MR\% decreases with increasing weight percentage of fCNT.  For all the temperatures (4.2K,10K,20K and 30K) measured, 10wt\% CNT/PANI composite shows negative magnetoresistance and value of the MR\% decreases with increasing temperature but same wt\% of fCNT/PANI composite shows the positive MR and increases with decreasing temperature(fig.~\ref{fig:final10}). %( ref. fig~\ref{fig:bipolaronfitted}).
\begin{figure}[tbh]
	\begin{center}
		\includegraphics[width=0.5\textwidth]{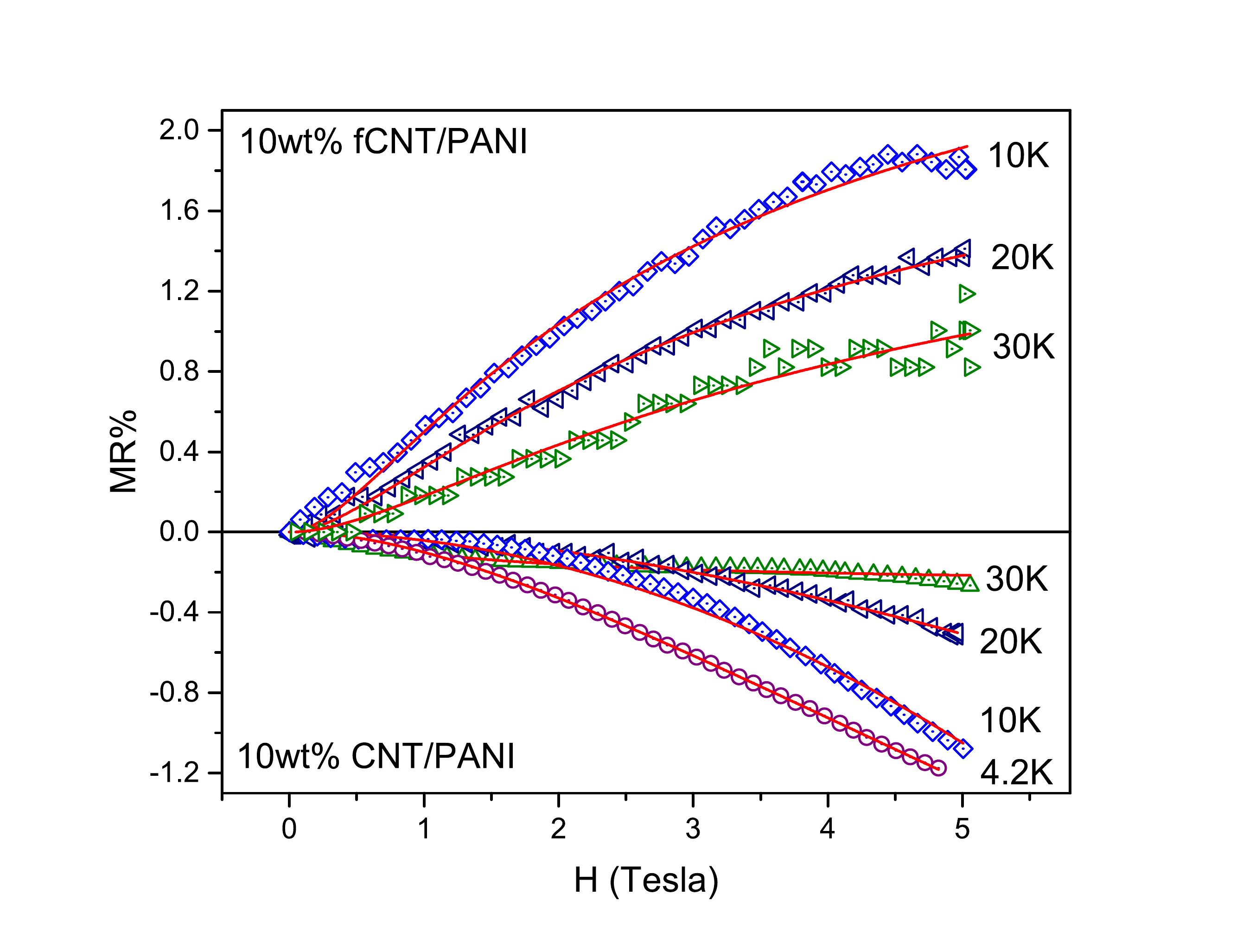}
		\small{\caption{ Plot of MR\% vs magnetic field for differnt temperature of 10wt\% CNT/PANI and fCNT/PANI fitted with bipolaron model.\label{fig:bipolaronfitted}}}
	\end{center}
\end{figure}

The transition of negative to positive MR due to functionalization of CNT can be explained by bipolaron model given by Bobbert et.al.~\cite{PhysRevLett.99.216801} In fig.~\ref{fig:bipolaronfitted} the negative MR data for all temperatures(4.2K,10K,20K and 30K) and positive MR data(except 4.2K) are well fitted with the empirical non-Lorentzian line shape equation
\begin{eqnarray}
MR(B) = MR_{\infty} \frac{B^2}{(|B|+B_{0})^2}
\end{eqnarray}
Where $MR_{\infty}$ is the MR value at infinite magnetic field(B), $B_{0}$ is the characteristic field width. The value of $B_{0}$ depends on the branching ratio
$b=r_{\alpha\longrightarrow\beta}/r_{\alpha\longrightarrow e}$, where $r_{\alpha\longrightarrow\beta}$ is the rate at which a polaron hops from a singly occupied site $\alpha$ into another neighbouring singly occupied site $\beta$ to form a bipolaron and $r_{\alpha\longrightarrow e}$ is the rate at which polaron hops from site $\alpha$ into the unoccupied site of environment bypassing the site $\beta$. 
We have seen that the $B_{0}$ value increases from 1.57T to 2.51T as temperature increases from 10K to 30K for fCNT/PANI composite. As the temperature increases the thermal energy of polarons increase which helps to hop from the site $\alpha$ into the longer distance empty sites i.e decreases the probability of bipolaron formation at site $\beta$ that's why the value of positive MR decreases and increases $B_{0}$ value.         

In this bipolaron model competition between two charge transport mechanism("spin blocking" and "Increase in polaron population with increasing magnetic field") along with parameter on-site Coulomb repulsive energy between two polarons and long range Coulomb repulsion, gives the transition from positive to negative MR. The charge carriers interacting with phonon creates polaron, align the spin with randomly oriented local hyperfine field created by hydrocarbon molecules in polymer. These polarons can form bipolaron with the cost of energy U(on-site coulomb repulsive energy of two polarons). The on-site exchange interaction favours only singlet bipolaron state but two polarons of same spin component along common quantization axis has zero probability to form singlet bipolaronic state, due to this spin blocking mechanism the resistivity increases with increasing magnetic field, gives positive MR. On the other hand by the increase of magnetic field the density of polaron increases at the expense of bipolaron, but the on-site Coulomb energy(U) is so large that hopping probability of polaron is reduced. Incorporating the long range coulomb repulsion(V) between the polarons decreases the effective on-site energy[(U-V), to form bipolaron] that favours to hop polaron to form bipolaron and enhances the conductivity, shows negative MR.
   
   Bipolaron model predicts  that there will be a transition of MR from negative to positive by decreasing the long-range Coulomb interaction between polarons for fixed temperature(T) and external applied electric field(E). The long-range and short-range(on-site) Coulomb energy has strong dependency on disorder and doping in the organic semiconductors.~\cite{PhysRevB.96.035104} Functionalization of CNT decreases the long-range Coulomb interaction compared to only CNT in the composite as discussed previously. This decrease of long-range interaction increases the effective on-site energy, reduces the probability to form bipolaron. That might be the reason to change sign of MR from negative to positive due to functionalization of CNT for all temperatures.

However, at very low temperature(4.2K) the thermal energy of charge carriers and number of phonon are so small that effect of disorder dominates in the charge transport mechanism. 
\begin{figure}[tbh]
	\begin{center}
		\includegraphics[width=0.5\textwidth]{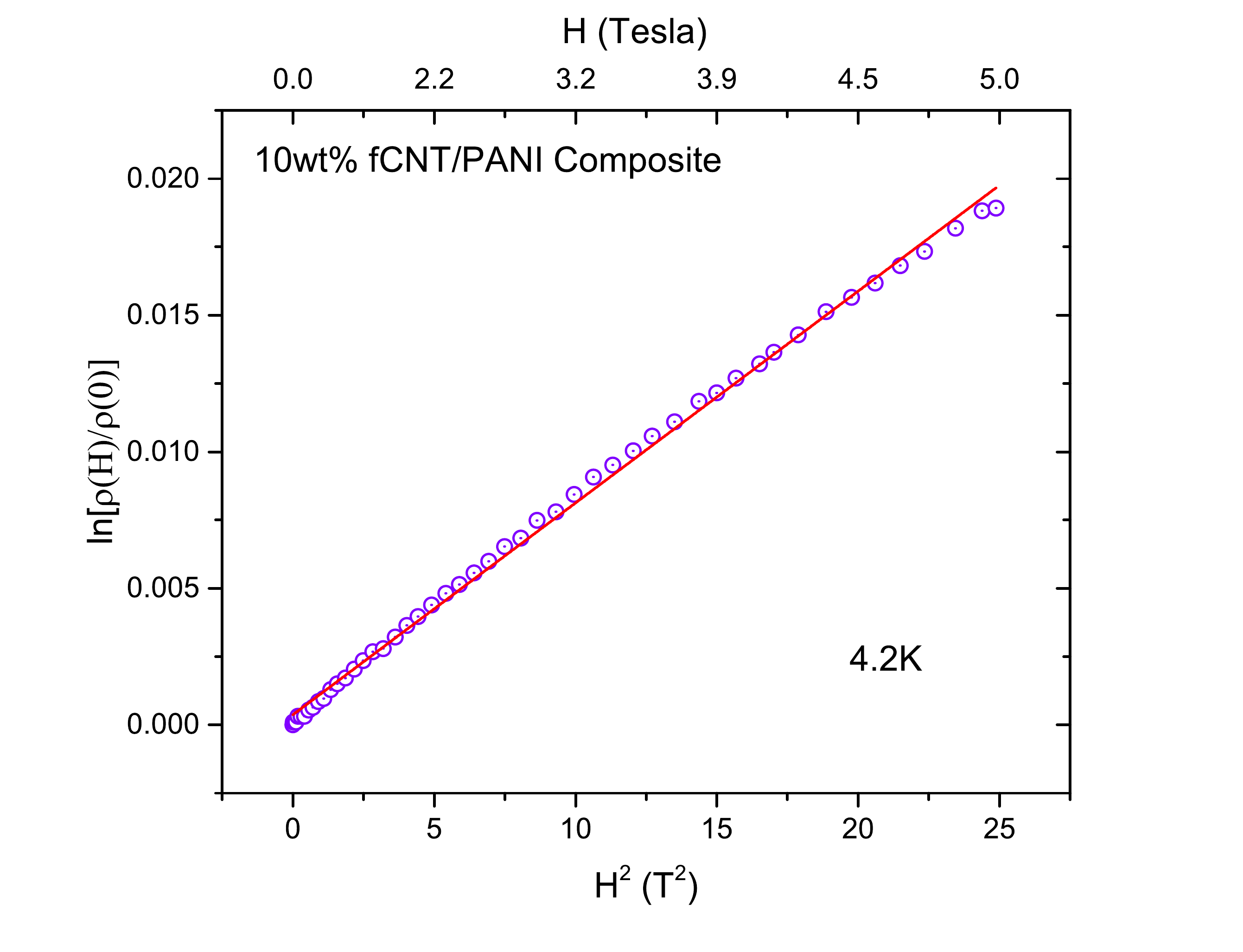}
		\small{\caption{(a) Plot of $ln[\rho(H)/\rho(0)]$ Vs $H^{2}$ for 10wt\% fCNT/PANI composite at 4.2K\label{fig:fcnt4}}}
	\end{center}
\end{figure}
Then positive MR can be explained by disorder induced wavefunction shrinkage model. This model tells that MR\% increases with increasing magnetic field due to shrinkage of hopping electron wavefunction. In strongly disorder system, in VRH region, weak field MR can be expressed as follows:~\cite{shklovskii2013electronic,rosenbaum2001magnetoresistance'}
\begin{eqnarray}
ln\dfrac{\rho(H)}{\rho(0)}  \approx t_2(\frac{e^2 a_0^4}{\hbar^2})(\frac{T_0}{T})^\frac{3}{4}H^2
\label{lfpmr}
\end{eqnarray}

 where $t_2=\frac{5}{2016}$, and $a_0$ is the Bohr radius, approximately equal to the localization length. The positive MR for 10wt\% fCNT/PANI is well fitted with VRH model $H^2$ law at 4.2K(fig.~\ref{fig:fcnt4}). We have extracted the localization length 8.23nm using the equation~\ref{lfpmr}.
  
   \section{Conclusion}
   We can conclude that functionalization of CNT and PANI composite shows huge variation(seven order) in resistivity which is much greater than same wt\% CNT/PANI composite at low temperature. The resistivity variation with temperature follows Mott VRH model and below a certain temperature it follows ES model. There is a transition of MR from negative to positive due to functionalization of CNT in 10wt\% composite with PANI. The fCNT contributes more disorder in the composites compared to CNT confirmed by Raman spectroscopy. The long-range Coulomb interaction is screened by disorder, increases the on-site Coulomb energy which hinders to form bipolaron. On the basis of `Bipolaron' model, decrease of long-range Coulomb interaction may be the cause of the MR transition from negative to positive. At 4.2K the positive MR of 10wt\% fCNT/PANI composite has explained by `wavefunction shrinkage' model.     
       
%\bibliographystyle{plain}

%\bibliography{bibfile}

%

\end{document}